\documentclass[aps,prc,twocolumn,nofootinbib,amsmath,amssymb, 10pt]{revtex4-2}
\usepackage[utf8]{inputenc}
\usepackage{amsmath}
\usepackage{graphicx}
\usepackage{float}
\usepackage{hyperref}
\usepackage{lipsum}
\usepackage{listings}
\usepackage{listingsutf8}
\usepackage{xcolor}
\usepackage{amssymb}
\usepackage{textcomp}
\usepackage{units}
\usepackage{bbm}
\usepackage{enumitem}  
\usepackage{url}
\usepackage[english]{isodate}
\usepackage{bm}
\usepackage{mdframed}
\usepackage[capitalise]{cleveref}
\usepackage{enumerate}
\usepackage{dsfont}
\usepackage{dcolumn}
\usepackage{physics}
\usepackage{color}
\usepackage{placeins}
\usepackage{slashed}
\usepackage{amssymb}
\usepackage{hyperref}
\usepackage[capitalise]{cleveref}
\usepackage[caption=false]{subfig}
\usepackage{braket}

\usepackage{mathrsfs}
\usepackage[normalem]{ulem}

\renewcommand{\Tr}{\operatorname{Tr}}

\newcommand*{\thetacm}{\theta_\mathrm{c.m.}}

\newcommand*{\NNLO}{N$^2$LO}

\newcommand*{\Tl}{T_\mathrm{lab}}

\usepackage{todonotes}

\begin{document}

\title{Entanglement and accidental symmetries in the nucleon-nucleon system}
\author{Alma L. Cavallin} 
\email{alma.cavallin@chalmers.se}
\author{Oliver Thim}
\author{Christian Forssén}
\affiliation{Department of Physics, Chalmers University of Technology, SE-412 96, Göteborg, Sweden}

\date{\today}
\noindent
\begin{abstract} 
We study the connection between accidental symmetries in the nuclear interaction and spin entanglement in two-nucleon scattering. Specifically, we incorporate different levels of Wigner $SU(4)$ and Serber symmetries into leading-order potentials derived from chiral effective field theory. We conduct a quantitative analysis by computing the full $S$ matrix, demonstrating that the neutron-proton spin entanglement can be related to the symmetry properties of the interaction and the presence of certain operators and partial waves. Furthermore, we study the order-by-order evolution of the spin entanglement, up to next-to-next-to-leading order in Weinberg power counting, for both neutron-proton and neutron-neutron scattering. Entanglement suppression is not observed in neutron-neutron scattering, which can be attributed to the Pauli principle and the absence of accidental symmetries in this system. We conclude that entanglement is a useful guide for studying the power counting and symmetries in nuclear interactions derived from effective field theories.

\end{abstract}

\maketitle

\section{Introduction}
\label{sec:introduction}
Nuclear forces derived from chiral effective field theory ($\chi$EFT) naturally incorporate the symmetries inherent in the underlying theory of quantum chromodynamics (QCD) \cite{Weinberg:1978kz, Machleidt:2011zz}. 
In contrast, the consequences of \emph{accidental} (emerging) nuclear symmetries, such as Wigner $SU(4)$ \cite{Wigner:1936dx} and Serber \cite{serber:1992} symmetries, often remain hidden. Being accidental symmetries, they 
might stem from the large $N_c$ expansion of QCD \cite{CalleCordon:2008cz, CalleCordon:2009ps} or from something completely different. \textcite{Beane:2018oxh} identified a striking connection between spin-entanglement suppression in $S$-wave neutron-proton ($np$) scattering and Wigner $SU(4)$ symmetry for (low-energy) nuclear forces. These authors conjectured that dynamical entanglement suppression gives rise to emergent symmetries and provides powerful constraints that predict the nature of strong interactions in the infrared domain. 

Wigner $SU(4)$, or the supermultiplet scheme, was suggested in 1937 by Wigner \cite{Wigner:1936dx} and Hund \cite{Hund:1937} independently of each other. Inspired by the charge independence hypothesis, Wigner proposed that treating the nuclear force as independent of the ordinary spin could serve as a useful additional approximation~\cite{Wigner:1936dx}. A Wigner $SU(4)$ invariant nuclear potential then only depends on the spatial coordinates and gives identical interactions in partial waves with the same orbital angular momentum \cite{CalleCordon:2008cz}. A nuclear potential expressed solely in terms of Wigner $SU(4)$ invariants is thus not expected to generate any spin entanglement.

An approximate Wigner $SU(4)$ symmetry in atomic nuclei has been observed in several contexts \cite{Parikh:1978nc, VanIsacker:1997vh, LiMuli:2025zro}. For the $np$ system system, the large $N_c$ expansion of QCD predicts a nearly Wigner-invariant interaction in even-parity partial waves \cite{Kaplan:1996rk}. Furthermore, pionless effective field theory is Wigner $SU(4)$ symmetric at leading order (LO) in the limit of infinite scattering lengths \cite{Mehen:1999qs}. The anomalously large scattering lengths of the $np$ system can therefore be regarded as an indicator of Wigner $SU(4)$ symmetry \cite{Mehen:1999qs, CalleCordon:2008cz}. Additionally, Arriola \emph{et al}. have suggested that Wigner $SU(4)$ is an approximately good \emph{long-distance symmetry} for even-parity partial waves, where they defined a long-distance symmetry as one that is only broken by (short-distance) counterterms in the potential~\cite{CalleCordon:2008cz, CalleCordon:2009ps}. 

The accidental Serber symmetry is the observation that the $np$ differential cross section is close to symmetric around $\theta_{\text{c.m.}}=90^\circ$ for low energies, where $\theta_{\text{c.m.}}$ is the scattering angle in the 
center of mass (c.m.) frame \cite{CalleCordon:2009ps, serber:1992, Christian:1952}. 
For neutron-neutron ($nn$) and proton-proton ($pp$) scattering, the differential cross section is inherently symmetric about $\theta_{\text{c.m.}}=90^\circ$ due to the indistinguishability of the particles \cite{Taylor72, Christian:1952}. 

Following the work of \textcite{Beane:2018oxh}, several further studies have investigated spin entanglement in few-nucleon systems. For example, spin entanglement was studied from a quantum-information perspective \cite{Low:2021ufv}, for $S$ waves using phenomenological models \cite{Kirchner:2023dvg, Bai:2022hfv}, for mixed spin entanglement \cite{Bai:2023}, and for the full $S$ matrix using the PWA93 model \cite{Bai:2023tey}.
Most of the work on entanglement in the two-nucleon sector has focused on $np$ scattering in the far infrared (only $S$ waves), either using phenomenological nuclear interactions or by analyzing empirical phase shifts. Spin entanglement in $\chi$EFT was only briefly examined in~\cite{Teng:2024exc}, using a power-counting scheme with perturbative pions. Additionally, $S$-wave spin entanglement in $nn$ scattering was discussed in~\cite{Hu:2025lua}. 

In this work, we perform a quantitative study of spin entanglement and accidental symmetries in the $np$ and $nn$ systems. We start by modifying the LO Weinberg power counting (WPC) \cite{Weinberg:1990rz, Weinberg:1991um} potential to incorporate Wigner $SU(4)$ and Serber symmetries to various degrees. For these modified potentials, we then study the spin entanglement generated by the full $S$ matrix for $np$ scattering as a function of the scattering energy and angle. This enables us to investigate the proposed connection between dynamical entanglement suppression and emergent symmetries---extending the analysis into kinematic regimes where the $S$-wave approximation is no longer valid.
Furthermore, we study the order-by-order evolution of spin entanglement in WPC up to \NNLO{} and compare the result with the phenomenological Nijmegen I (NijmI) potential, for both $np$ and $nn$ scattering.

The paper is organized as follows. In Section \ref{sec:formalism}, we present the theoretical formalism, including metrics and methods for computations of spin entanglement in $np$ and $nn$ scattering. We also describe our construction of symmetry-enhanced versions of the LO WPC potential. In Section \ref{sec:entanglement_results}, the numerical results for spin entanglement in $np$ and $nn$ scattering are presented and discussed. On a more technical level, we examine the differences in spin entanglement between forward and nonforward scattering directions, paying particular attention to the case of isotropic $S$ waves, where the distinction becomes subtle. Finally, we summarize and conclude the study in Section \ref{sec:conclusion}. 

\section{Formalism}
\label{sec:formalism}
The two-nucleon scattering operator, $S$, can be decomposed as \cite{Taylor72, Glockle}
\begin{equation}
    \begin{split}
        S(\boldsymbol{p}',\boldsymbol{p}) = \frac{\delta(p'-p)}{p^2}[\delta^{(2)}(\hat{\boldsymbol{p}}'-\hat{\boldsymbol{p}})\mathbbm{1} + \frac{ip}{2\pi}\boldsymbol{M}(\boldsymbol{p}',\boldsymbol{p})],
        \label{eq:scattering_operator}
    \end{split}
\end{equation}
where $\boldsymbol{p}$ ($\boldsymbol{p}'$) is the incoming (outgoing) relative momentum in the c.m.~frame and the modulus and direction of the momentum vectors are denoted by $p$ and $\hat{\boldsymbol{p}}$, respectively. Both the identity matrix $\mathbbm{1}$ and the $M$ matrix $\boldsymbol{M}(\boldsymbol{p}', \boldsymbol{p})$ are $4\times4$ matrices in the two-nucleon spin space $\mathscr{H}_{\text{spin}}$. 
The nontrivial part of $S$ is proportional to $\boldsymbol{M}(\boldsymbol{p}',\boldsymbol{p})$ which contains all dynamical information about the scattering process \cite{Glockle}. 
Due to rotational invariance, there is no azimuthal dependence and $\boldsymbol{M}(p, \theta_{\text{c.m.}}) \equiv \boldsymbol{M}(\boldsymbol{p}',\boldsymbol{p})$, where $\theta_{\text{c.m.}}$ is the angle between $\boldsymbol{p}$ and $\boldsymbol{p}'$ in the scattering plane. For brevity, we will occasionally omit the dependence of $\boldsymbol{M}$ on $p$ and $\theta_{\text{c.m.}}$. 

For nonforward scattering directions ($\theta_{\text{c.m.}}\neq 0^\circ$), the normalized outgoing spin state of the scattering process $\ket{\chi_{\text{out}}}\in \mathscr{H}_{\text{spin}}$ is given by \cite{Nielsen:2010}
\begin{equation}
    \ket{\chi_{\text{out}}} = \frac{\boldsymbol{M}\ket{\chi_{\text{in}}}}{\sqrt{\braket{\chi_{\text{in}}|\boldsymbol{M}^\dagger\boldsymbol{M}|\chi_{\text{in}}}}},
    \label{eq:chi_out}
\end{equation}
where $\ket{\chi_{\text{in}}}\in\mathscr{H}_{\text{spin}}$ is the initial spin state. 
The trivial part of the scattering operator, see Eq. \eqref{eq:scattering_operator}, only affects the outgoing state in the forward direction ($\theta_{\text{c.m.}}=0^\circ$). In this work, we mainly study entanglement in nonforward directions. However, we also discuss entanglement in the forward direction $\theta_{\text{c.m.}}=0^\circ$ and its relation to the $S$-wave approximation used in previous studies.

\subsection{Entanglement measures}
\label{subsec:entanglement_measures}
Here, we will first discuss entanglement measures for distinguishable particles, as used in $np$ scattering. We will then show how to treat spin entanglement of identical particles, which is relevant for the case of $nn$ scattering.

Let us first consider two distinguishable nucleons for which the two-nucleon spin state is assumed to be pure\footnote{We only consider pure-state and not mixed-state entanglement. For the latter, the situation is more complicated. In fact, for a mixed state, there is no straightforward way to quantify the entanglement of a given density matrix, in contrast to pure states where the entanglement of the state can be completely quantified using an appropriate pure-state entanglement measure \cite{Bennett:1996gf}.}. We start from an unentangled initial state. For distinguishable particles, a pure bipartite state is unentangled if and only if it can be written as a tensor-product state \cite{Horodecki:2009zz, Ghirardi:2004}. Therefore, the initial two-nucleon spin state is given by 
\begin{equation}
    \ket{\chi_{\text{in}}} = \ket{\chi_1}\otimes\ket{\chi_2},
    \label{eq:tensor_product_state}
\end{equation}
where 1 and 2 stand for nucleon 1 and 2, respectively. The final spin state $\ket{\chi_{\text{out}}}$ is then also a pure state, but possibly entangled. The separability criterion \eqref{eq:tensor_product_state} can be used to formulate entanglement measures for pure states, such as the von Neumann entropy of the reduced density matrix \cite{Horodecki:2009zz, Plenio:2007zz, Wootters:1997id}. We follow Ref.~\cite{Beane:2018oxh} and use the two-entropy $\mathcal{E}_{\text{2E}}(\ket{\chi_{\text{out}}})$, which is the leading term in the Taylor expansion of the von Neumann entropy. The two-entropy is given by \cite{Low:2021ufv}
\begin{equation}
    \mathcal{E}_{\text{2E}}(\ket{\chi_{\text{out}}}) = 1 - \Tr_1\rho_1^2,
    \label{eq:two_entropy}
\end{equation}
where $\rho_1 = \Tr_2\ketbra{\chi_{\text{out}}}{\chi_{\text{out}}}$ is the reduced density matrix. As for any entanglement measure, the two-entropy does not increase under local operations and classical communication \cite{Horodecki:2009zz, Bennett:1996gf} from which it follows that it attains its minimal (maximal) value for unentangled (maximally entangled) states \cite{Plenio:2007zz}. The range of the two-entropy is $0\leq \mathcal{E}_{\text{2E}} \leq 1/2$ \cite{Low:2021ufv}.

The entanglement of $\ket{\chi_{\text{out}}}\,\propto\,\boldsymbol{M}\ket{\chi_{\text{in}}}$ will in general depend on the initial state $\ket{\chi_{\text{in}}}$\footnote{In some special cases the entanglement will not depend on the initial spin state, for example, when $\boldsymbol{M}$ is proportional to the identity matrix (if the initial state is unentangled) or a Bell projector.}. It is therefore useful to define a state-independent entanglement measure for the operator $\boldsymbol{M}$. The so-called \emph{entanglement power}, introduced in Ref.~\cite{Zanardi:2001zza}, quantifies the average entanglement produced by an operator when acting on unentangled states. The entanglement power of the $M$ matrix, $\mathcal{E}_{\text{EP}}(\boldsymbol{M})$, based on the two-entropy reads
\begin{equation}
    \mathcal{E}_{\text{EP}}(\boldsymbol{M}) \equiv \overline{\mathcal{E}_{\text{2E}}(\boldsymbol{M}\ket{\chi_{\text{in}}})},
    \label{eq:entanglement_power}
\end{equation}
where the bar denotes the average over all unentangled tensor-product states $\ket{\chi_{\text{in}}}$. Of course, other pure-state entanglement measures can be used as a basis for the entanglement power. In Ref.~\cite{Kirchner:2023dvg}, different orders of the Taylor-expanded von Neumann entropy and Rényi entropies were compared, and it was shown that the different entropies are equally informative. 

We now turn to $nn$ scattering, where the two nucleons are identical.
The total wave function for identical fermions is antisymmetric in the interaction region. 
However, in a scattering process it is the asymptotic \emph{free} states that are measurable, and the interaction itself, which is encoded in the scattering operator, can never be observed. Since the asymptotic states are noninteracting and the nucleons effectively can be treated as distinguishable there is no need to antisymmetrize the two-nucleon wave function in the in- and out-asymptotes. The indistinguishability of the particles only has to be incorporated into the scattering operator \cite{Taylor72}. 

This \emph{effective distinguishability} makes it possible to apply entanglement measures for distinguishable particles, even for $nn$ scattering \cite{De_Muynck:1975, Herbut:1987, Herbut:2006, Tichy:2012, Cunden:2014, Benatti:2020xgb}. The concerns about entanglement for identical particle systems, see e.g., Refs. \cite{Benatti:2014gaa, Ghirardi:2002, Benatti:2020xgb, Ghirardi:2004, Eckert:2002inq}, should only be taken into account when we are considering entanglement in systems where the identical constituents have overlapping spatial wave functions (and hence cannot be effectively distinguished). For example, in the study of the entanglement between constituents in atomic nuclei \cite{Tews:2022yfb, Robin:2020aeh}.

\subsection{Symmetry-enhanced leading order potentials for the neutron-proton system}
\label{subsec:symmetry_potentials}
The Wigner $SU(4)$ spin-isospin symmetry does not discriminate between channels of different spin-isospin configurations but the same spatial configuration. The $np$ channels can thus be sorted into supermultiplets 
\begin{equation}
    ({^1\!S_0},{^3\!S_1}), \quad ({^1\!{P}_1}, {^3\!{P}_{0,1,2}}), \quad ({^1\!{D}_2}, {^3\!{D}_{1,2,3}}), \quad ...,
    \label{eq:supermultiplets}
\end{equation}
such that partial waves belonging to the same supermultiplet have identical interactions for a Wigner $SU(4)$ invariant potential \cite{CalleCordon:2008cz}. The spectroscopic notation $^{2S+1}\!L_J$ is used in \cref{eq:supermultiplets}, where $S$ is the coupled spin, $L$ is the orbital angular momentum, and $J$ is the total angular momentum. 

In contrast to the Wigner $SU(4)$ symmetry, the Serber symmetry is not a continuous group but the observation that the $np$ differential cross section is approximately symmetric around $\theta_{\text{c.m.}}=90^\circ$ at low scattering energies. Exact symmetry around $90^\circ$ can be obtained by setting the interaction to zero in all odd-parity partial waves \cite{Christian:1952}. We could also get symmetry around $\theta_{\text{c.m.}}=90^\circ$ in other ways, for example by only having spin-singlets in even-parity partial waves and spin-triplets in odd-parity partial waves, which is the situation for two identical spin 1/2 fermions.

We want to study the consequences of incorporating the Wigner $SU(4)$ and the Serber symmetries to different degrees into the LO WPC potential $V_{\text{WPC,LO}}$. This potential is derived from the LO Feynman diagrams, which in WPC are a nonderivative contact and the static one-pion exchange (OPE). The contact part contains low-energy constants (LECs) which parametrize the short-distance dynamics and must be fitted to data. The OPE diagram gives the LO description of the long-distance nuclear force \cite{Machleidt:2011zz}. 

The LO WPC potential in momentum space reads \cite{Machleidt:2011zz}
\begin{equation}
        V_{\text{WPC,LO}}(\boldsymbol{q}) =C_S + C_T\boldsymbol{\sigma}_1\cdot\boldsymbol{\sigma}_2 + V_{\text{OPE}}(\boldsymbol{q}),
    \label{eq:LO_WPC_potential_momentum_space}
\end{equation} 
where
\begin{equation}
V_{\text{OPE}}(\boldsymbol{q}) =  -\frac{g_A^2}{4f_\pi^2}\frac{\boldsymbol{\sigma}_1\cdot\boldsymbol{q}\boldsymbol{\sigma}_2\cdot\boldsymbol{q}}{\boldsymbol{q}^2 + m_\pi^2}\boldsymbol{\tau}_1\cdot\boldsymbol{\tau}_2,
    \label{eq:OPE_potential_momentum_space}
\end{equation} 
and $\boldsymbol{q}=\boldsymbol{p}'-\boldsymbol{p}$ is the momentum transfer. Furthermore, $g_A$ is the axial coupling constant, $f_\pi$ is the pion decay constant, $m_\pi$ is the average pion mass and $\boldsymbol{\sigma}_i$ ($\boldsymbol{\tau}_i$) is a vector of Pauli matrices in spin (isospin) space for nucleon $i$. The contact terms only contribute to $S$ waves ($L = 0$). For $np$ scattering, there are two $S$-wave channels and the partial-wave projected contact terms read \cite{Machleidt:2011zz}
\begin{equation}
    \tilde{C}_{{^1\!S_0}} = 4\pi(C_S - 3C_T), \quad \text{and} \quad \tilde{C}_{{^3\!S_1}} = 4\pi(C_S + C_T).
\end{equation}
Wigner $SU(4)$ symmetry in the contact part of the LO WPC potential implies $\tilde{C}_{{^1\!S_0}} = \tilde{C}_{{^3\!S_1}}$, or equivalently $C_T=0$. Accordingly, what we refer to as short-distance Wigner $SU(4)$ symmetry can be imposed by assigning equal values to the contact terms in the two $S$-wave channels, which we then denote $\tilde{C}_{\text{W}}$.

The only invariants of Wigner $SU(4)$ are the identity operator $\mathbbm{1}$ and the Majorana exchange operator $P_M$, which exchanges the spatial coordinates between the two nucleons \cite{Wigner:1936dx}. Due to the Pauli principle $P_M = -P_\sigma P_\tau$, where $P_\sigma$ ($P_\tau$) is the spin (isospin) exchange operator \cite{Brink:1965}. The explicit form of $P_M$ is
\begin{equation}
   P_M = -\frac{1}{4}(\mathbbm{1} + \boldsymbol{\sigma}_1\cdot\boldsymbol{\sigma}_2)(\mathbbm{1}+\boldsymbol{\tau}_1\cdot\boldsymbol{\tau}_2).
\end{equation}

Ref.~\cite{CalleCordon:2008cz} suggests that Wigner $SU(4)$ symmetry is strongly broken at short distances and in odd-parity partial waves, while it is an approximate symmetry for even-parity partial waves at long distances. Short-distance symmetry breaking is due to different $np$ scattering lengths in the two $S$-wave channels and is expected to be notable for momenta $p< 1/a_S$, where $a_S$ is the scale of the scattering lengths. The energy window in which Wigner $SU(4)$ symmetry is applicable is thus enlarged by the anomalously large $np$ scattering lengths. For even-parity partial waves there exist two more invariants, besides $\mathbbm{1}$ and $P_M$, namely
\begin{equation}
    \boldsymbol{\sigma}_1\cdot\boldsymbol{\sigma}_2\boldsymbol{\tau}_1\cdot\boldsymbol{\tau}_2 \quad  \text{and} \quad \boldsymbol{\sigma}_1\cdot\boldsymbol{\sigma}_2 + \boldsymbol{\tau}_1\cdot\boldsymbol{\tau}_2.
    \label{eq:even_parity_invariants}
\end{equation}
This can be seen by computing the expectation values of these operators with respect to a two-nucleon state $\ket{ST}$, where $S$ ($T$) denotes the coupled spin (isospin). 

The full OPE potential breaks Wigner $SU(4)$ symmetry. Long-distance Wigner $SU(4)$ symmetry can nevertheless be incorporated by identifying which parts of the OPE potential are Wigner symmetric for even-parity partial waves. For this purpose, it is convenient to transform the OPE potential~\eqref{eq:OPE_potential_momentum_space} to position space
\begin{equation}
    \begin{split}
        V_{\text{OPE}}(\boldsymbol{x}) = &\frac{g_A^2m_\pi^2}{16\pi f_\pi^2}\frac{e^{-m_\pi r}}{r}[\frac{1}{3}\boldsymbol{\sigma}_1\cdot\boldsymbol{\sigma}_2\\
        &+ S_{12}(\hat{\boldsymbol{x}})(\frac{1}{m_\pi^2r^2} + \frac{1}{m_\pi r} + \frac{1}{3})]\boldsymbol{\tau}_1\cdot\boldsymbol{\tau}_2,
    \end{split}
    \label{eq:OPE_potential_position_space}
\end{equation}
where $\boldsymbol{x}$ is the relative coordinate, $r= |\boldsymbol{x}|$, $\hat{\boldsymbol{x}} = \boldsymbol{x}/r$ and the tensor operator $S_{12}(\hat{\boldsymbol{x}})$ is given by \cite{Brink:1965}
\begin{equation}
    S_{12}(\hat{\boldsymbol{x}}) = 3\boldsymbol{\sigma}_1\cdot\hat{\boldsymbol{x}}\boldsymbol{\sigma}_2\cdot\hat{\boldsymbol{x}} - \boldsymbol{\sigma}_1\cdot\boldsymbol{\sigma}_2.
    \label{eq:tensor_operator}
\end{equation}
The tensor operator mixes partial waves with the same $J$ but different $L$ and is zero for spin-singlet waves \cite{Brink:1965}. Thus, the expectation value of $S_{12}(\hat{\boldsymbol{x}})$ can vary among partial waves with the same $L$, implying that this operator breaks Wigner $SU(4)$ symmetry. The non-tensor part of the OPE potential has the spin-isospin operator structure $\boldsymbol{\sigma}_1\cdot\boldsymbol{\sigma}_2\boldsymbol{\tau}_1\cdot\boldsymbol{\tau}_2$ which we recognize as one of the even-parity invariants from \cref{eq:even_parity_invariants}. The even-parity Wigner-symmetric part of the OPE potential can thus be extracted by removing the tensor operator term in \cref{eq:OPE_potential_position_space}. We call this potential $W(\boldsymbol{x})$. In momentum space it reads
\begin{equation}
    W(\boldsymbol{q}) = \frac{g_A^2m_\pi^2}{12f_\pi^2}\boldsymbol{\sigma}_1\cdot\boldsymbol{\sigma}_2\boldsymbol{\tau}_1\cdot\boldsymbol{\tau}_2\frac{1}{\boldsymbol{q}^2 + m_\pi^2}.
    \label{eq:Wigner_symmetric_OPE_momentum_space}
\end{equation}

Using the above considerations, five modified versions of the LO WPC potential with different symmetry properties at short and long distances were constructed. They are presented in \cref{tab:modified_LO_potentials_np}. 

\begingroup
\renewcommand{\arraystretch}{1.7}
\begin{table*}[!ht]
    \centering
    \caption{The LO WPC potential and five modified versions for the $np$ system. The subscripts refer to the symmetry properties. For potentials labeled with two subscripts, the first indicates whether Wigner symmetry is preserved ('sym') or broken ('break') in the $S$-wave contact potential, while the second denotes whether Serber symmetry is preserved or broken in odd-parity partial waves. All modified potentials, $V^{(1)}$--$V^{(5)}$, are Wigner symmetric at long distances in even-parity partial waves. The two potentials $V^{(1)}$ and $V^{(2)}$ are Serber symmetric in odd-parity partial waves, while $V^{(3)}$--$V^{(5)}$ are Serber breaking. The potential $V^{(5)}_{\text{no tensor}}$ is the LO WPC potential with the $S_{12}(\hat{\boldsymbol{x}})$ term removed.}
    \begin{tabular}{c c c c c c c} \hline\hline
         Potential  &$V_{\text{WPC,LO}}(\boldsymbol{q})$ &$V^{(5)}_{\text{no tensor}}(\boldsymbol{q})$ &$V^{(4)}_{\text{break,break}}(\boldsymbol{q})$ &$V^{(3)}_{\text{sym,break}}(\boldsymbol{q})$ &$V^{(2)}_{\text{break,sym}}(\boldsymbol{q})$ &$V^{(1)}_{\text{sym,sym}}(\boldsymbol{q})$      \\ \hline
         Even $L$ (short) &$\tilde{C}_{{^1\!S_0}}$, $\tilde{C}_{{^3\!S_1}}$ &$\tilde{C}_{{^1\!S_0}}$, $\tilde{C}_{{^3\!S_1}}$  &$\tilde{C}_{{^1\!S_0}}$, $\tilde{C}_{{^3\!S_1}}$ &$\tilde{C}_{\text{W}}$ &$\tilde{C}_{{^1\!S_0}}$, $\tilde{C}_{{^3\!S_1}}$ &$\tilde{C}_{\text{W}}$ \\
         Even $L$ (long) &$V_{\text{OPE}}(\boldsymbol{q})$ &$W(\boldsymbol{q})$ &$W(\boldsymbol{q})$ &$W(\boldsymbol{q})$ &$W(\boldsymbol{q})$ &$W(\boldsymbol{q})$  \\
         Odd $L$ &$V_{\text{OPE}}(\boldsymbol{q})$ &$W(\boldsymbol{q})$ &$V_{\text{OPE}}(\boldsymbol{q})$ &$V_{\text{OPE}}(\boldsymbol{q})$ &0 &0      \\ \hline\hline 
    \end{tabular}
    \label{tab:modified_LO_potentials_np}
\end{table*}
\endgroup

\subsection{Numerical implementation}
\label{subsec:numerical_impl}
The $M$-matrix elements $M^S_{M_S'M_S}(p, \theta_{\text{c.m.}})$, with coupled spin $S$, and initial (final) spin projection $M_S$ ($M'_S$), are computed as \cite{Glockle} 
\begin{equation}
    \begin{split}
        M_{M_S'M_S}^S(p, \theta_{\text{c.m.}}) =&\frac{\sqrt{\pi}}{ip}\sum_{LL'J}i^{L-L'}Y_{L'M_S-M_S'}(\theta_{\text{c.m.}})\\
        &\times(2J+1)\sqrt{2L+1}\\
        &\times\begin{pmatrix}
            L'&S&J\\
            M_S-M_S'&M_S'&-M_S
        \end{pmatrix}\\
        &\times\begin{pmatrix}
            L&S&J\\
            0&M_S&-M_S
        \end{pmatrix}\left(S^{JST}_{L'L}(p)-\delta_{L'L}\right),
    \end{split}
    \label{eq:M_matrix_element}
\end{equation}
where $L$ ($L'$) is the initial (final) coupled orbital angular momentum and $Y_{LM_L}(\theta_{\text{c.m.}})$ denotes the spherical harmonics. As before, $J$ is the coupled total angular momentum and $T$ is the coupled isospin. The partial wave $S$-matrix elements are denoted $S^{JST}_{L'L}(p)$ where the super (sub) scripts denote conserved (non-conserved) quantum numbers under the assumption of isospin conservation \cite{Glockle}.
$S^{JST}_{L'L}(p)$ is computed by decomposing the Lippmann-Schwinger (LS) equation into partial waves following Ref.\ \cite{Erkelenz:1971caz} and solving the resulting matrix equations with matrix inversion, see Ref.\ \cite{Landau:1996}. Wigner 3j-symbols that enter the expression \eqref{eq:M_matrix_element} are calculated with the \textsc{WIGXJPF} code \cite{pywig}. 
Finally, we truncate the sum over $J$ in Eq. \eqref{eq:M_matrix_element} at $J=50$.
We use pion masses from the Particle Data Group \cite{PDG:2018}, nucleon masses from CODATA 2022 \cite{NIST:2022} and $f_\pi =92.4$ MeV \cite{Epelbaum:2004fk} in all potentials. The empirical value $g_A=1.276$ \cite{Liu:2010gA} is used in the LO potentials. 

Since the LS equation contains divergent integrals, the potentials must be regulated. For all potentials $V(\boldsymbol{p}',\boldsymbol{p})$, the following regulator is used 
\begin{equation}
    V(\boldsymbol{p}', \boldsymbol{p}) \rightarrow e^{-p'^6/\Lambda^6}V(\boldsymbol{p}',\boldsymbol{p})e^{-p^6/\Lambda^6},
\end{equation}
with $\Lambda=500$ MeV for the momentum cutoff.   

\begin{figure}[ht]
    \centering
    \includegraphics[width=\linewidth]{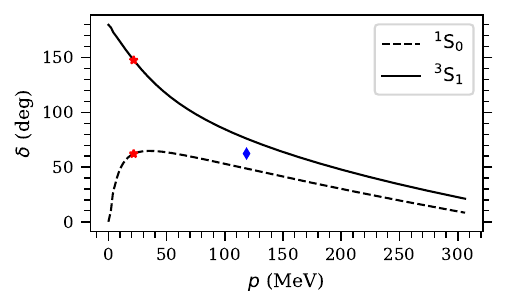}
    \caption{NijmI \cite{Stoks:1994wp} $S$-wave phase shifts. The red stars are the ${^1\!S_0}$ and ${^3\!S_1}$ phase shifts at $p = 22$ MeV ($T_{\text{lab}}=1$ MeV). The blue diamond is the average $S$-wave phase shift at $p = 119$ MeV ($T_{\text{lab}}=30$ MeV).}
    \label{fig:NijmI_S_phase_shifts}
\end{figure}

\label{subsubsec:LECs}
For the modified LO potentials in \cref{tab:modified_LO_potentials_np}, we tune the LECs by doing a least squares fit to single NijmI \cite{Stoks:1994wp} phase shifts collected from NN-OnLine \cite{nn-on-line}. For the potentials with a single LEC, the average $S$-wave phase shift at $T_{\text{lab}}=30$ MeV is used while for the potentials with two LECs, the $S$-wave phase shifts at $T_{\text{lab}}=1$ MeV are used, see \cref{fig:NijmI_S_phase_shifts}. The kinetic energy of the projectile particle in the laboratory frame $T_{\text{lab}}$ is related to the modulus of the relative momentum in the c.m.~frame $p$. We use the relativistic relation in all calculations.
Note that having a single LEC is unphysical in the sense that the ${^3\!S_1}-{^3\!{D}_1}$ channel contains the deuteron bound state while the uncoupled ${^1\!S_0}$ channel hosts a virtual (unbound) state. Still, these potentials are useful for studying the connection between short-distance Wigner symmetry and spin entanglement. 

We will also consider higher orders in WPC. Leading two-pion exchange diagrams appear at NLO as well as contact terms with derivatives. At N$^2$LO, further two-pion exchange diagrams enter. 
We use calibrated LECs from Ref. \cite{Carlsson:2015vda} in the WPC potentials and therefore follow their conventions.

To compute the entanglement power in \cref{eq:entanglement_power}, we represent the initial nucleon spin states using Bloch-sphere coordinates \cite{Nielsen:2010} and evaluate the resulting four-dimensional integral via Monte Carlo integration \cite{Press:2007ipz}. We use 1000 sampling points which results in an estimated relative error of 2.5\%.

\section{Entanglement results}
\label{sec:entanglement_results}
We now study the entanglement power as a function of the relative momentum, $p$, and the scattering angle, $\theta_{\text{c.m.}}$, in different scenarios. 

\subsection{Enhanced symmetries in the LO potential}
\begin{figure*}[htb]
    \centering
    \includegraphics[width=.9\linewidth]{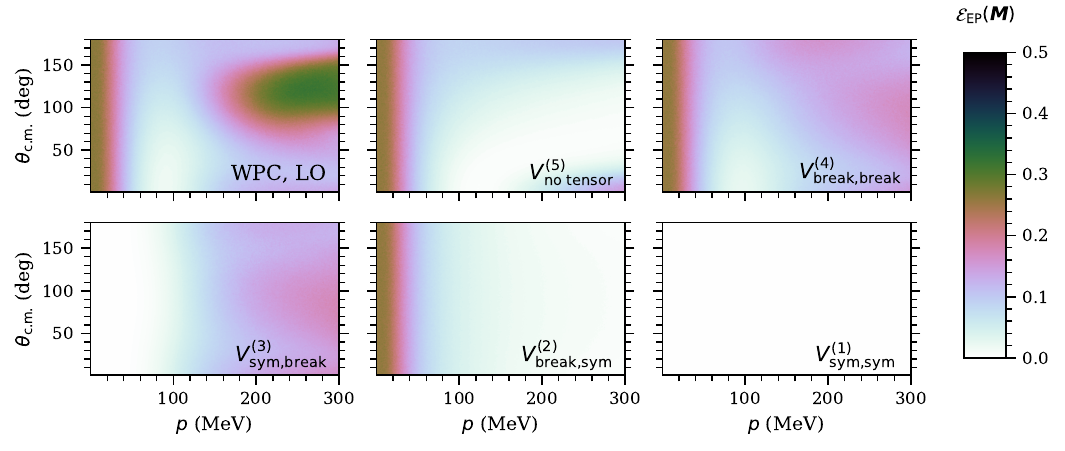}
    \caption{Entanglement power of the $M$ matrix $\mathcal{E}_{\text{EP}}(\boldsymbol{M})$ for the $np$ system using the LO WPC and the modified LO WPC potentials listed in \cref{tab:modified_LO_potentials_np}. Each LEC is tuned to a single $S$-wave phase shift, see Section \ref{subsubsec:LECs}, in all potentials (even the LO WPC potential).  $\mathcal{E}_{\text{EP}}(\boldsymbol{M})$ is evaluated for scattering angles $\theta_{\text{c.m}}\in[1^\circ, 180^\circ]$ and relative momenta $p\in [0.01,300]$ MeV.
    \label{fig:EP_modified_LO_np}}
\end{figure*}

We first consider the $np$ system described using the LO potential and its symmetry-enhanced variants. \Cref{fig:EP_modified_LO_np} presents the entanglement power for $\boldsymbol{M}(p,\thetacm)$ in the nonforward scattering directions for the different LO potentials listed in \cref{tab:modified_LO_potentials_np}. 

The upper left panel shows the result for the LO WPC potential. Areas of large entanglement power can be observed for low energies ($p \lesssim 50$~MeV) and for near-backwards scattering at higher energies ($p\gtrsim150$~MeV,  $100^\circ\lesssim\thetacm\lesssim150^\circ$). 

We begin by examining the impact of the Wigner-breaking tensor force on the entanglement power. 
The upper right panel shows the entanglement power for $V^{(4)}_{\text{break,break}}$, in which the tensor force is removed in the even-parity partial waves. This enhancement of Wigner $SU(4)$ symmetry at long distances makes the area of large entanglement power for $p \gtrsim 150$~MeV vanish. A moderate entanglement power remains for all angles and for $p \gtrsim 150$~MeV. Furthermore, the upper middle panel shows the entanglement power for $V^{(5)}_{\text{no tensor}}$, in which the tensor force is completely removed in all partial waves.
Now, the entanglement power for $p \gtrsim 150$~MeV is highly suppressed. We therefore conclude that the Wigner-breaking tensor force is the main source of entanglement in the kinematical region $p \gtrsim 150$~MeV. 

Next, we turn our attention to low energies, where significant entanglement power persists also after removing the tensor force.
Entanglement power in  the low energy region ($p \lesssim 50$~MeV) is present for all potentials in \cref{fig:EP_modified_LO_np}, except for $V^{(3)}_{\text{sym,break}}$ and $V^{(1)}_{\text{sym,sym}}$. These two potentials share the common feature of implementing  (i) even-parity long-distance Wigner symmetry and (ii) short-distance Wigner symmetry by having the same $S$-wave contacts in the ${^1\!S_0}$ and ${^3\!S_1}$ partial waves. The vanishing low-energy entanglement power can be explained by considering \cref{eq:M_matrix_element} for low energies using the effective range expansion \cite{Bethe:1949yr,Bai:2023tey}, which gives
\begin{equation}
    \lim_{p\rightarrow0}\boldsymbol{M} = -\frac{1}{2}(a_{{^1\!S_0}}+a_{{^3\!S_1}})\mathbbm{1} + \frac{1}{2}(a_{{^1\!S_0}}-a_{{^3\!S_1}})P_\sigma.
    \label{eq:very_low_energy_M_matrix}
\end{equation}
Here, $a_{{^1\!S_0}}$ and $a_{{^3\!S_1}}$ denote the ${^1\!S_0}$ and ${^3\!S_1}$ scattering lengths, respectively. Since the singlet and triplet scattering lengths are equal for $V^{(2)}_{\text{sym,break}}$ and $V^{(1)}_{\text{sym,sym}}$, \cref{eq:very_low_energy_M_matrix} forces the $M$ matrix to be proportional to the identity at low energies---which generates no entanglement.

The fully Wigner- and Serber-symmetric interaction $V^{(1)}_{\text{sym,sym}}$ gives zero entanglement power for all momenta and scattering angles. In fact, the $M$ matrix resulting from this potential is proportional to the identity operator in the entire kinematical region. 
Note that we would also get zero entanglement power if the $M$ matrix were proportional to $P_\sigma$, however, this does not correspond to Wigner $SU(4)$ symmetry.

Let us now turn to the mechanisms that enhance entanglement at low energies. The potential $V^{(2)}_{\text{break,sym}}$ breaks Wigner $SU(4)$ symmetry in the contact interaction only. From the lower middle panel of \cref{fig:EP_modified_LO_np}, we can therefore conclude that Wigner-breaking in the contact potential results in significant entanglement power solely at low energies. It might seem counterintuitive that this \emph{short-distance} Wigner breaking results in large entanglement power at \emph{low energies}.
Although the LECs parameterize the short-distance (or high-energy) dynamics---not explicitly resolved in the EFT---the two LO $S$-wave LECs are tuned to reproduce low-energy data. In $V^{(2)}_{\text{break,sym}}$ these LECs are calibrated using single-energy phase shifts at $\Tl=1$~MeV yielding the $S$-wave scattering lengths that enter \cref{eq:very_low_energy_M_matrix}. 
The small changes in the potential required to fine-tune the anomalously large $S$-wave scattering lengths have a vanishingly small effect at higher scattering energies \cite{Braaten:2004rn, RuizArriola}. This fine-tuning thus explains the absence of entanglement in $V^{(2)}_{\text{break,sym}}$ for $p \gtrsim 50$~MeV. 

The entanglement power for $V^{(3)}_{\text{sym,break}}$ and $V^{(4)}_{\text{break,break}}$ in \cref{fig:EP_modified_LO_np} reveal that Serber-breaking in odd-parity partial waves generates moderate entanglement power for momenta $p\gtrsim 150$ MeV. 
An additional modified version of the LO WPC potential was implemented---being identical to $V^{(2)}_{\text{break,sym}}$ apart from adding the OPE interaction in odd-parity spin-singlet waves. The resulting entanglement pattern (not included in \cref{fig:EP_modified_LO_np}) closely resembles that of $V^{(5)}_{\text{no tensor}}$. This observation suggests that the non-zero entanglement power near the forward and backward scattering directions in $V^{(5)}_{\text{no tensor}}$ arises from Serber symmetry breaking in odd-parity spin-singlet channels.

Finally, we observe that all potentials shown in \cref{fig:EP_modified_LO_np} exhibit vanishingly small entanglement power at $50 \text{ MeV } \lesssim p \lesssim150$~MeV, consistent with earlier studies \cite{Bai:2022hfv} of $np$ scattering. This observation is further explored in the next section by extending the analysis to $nn$ scattering and higher orders in the chiral expansion.

\subsection{Neutron-proton and neutron-neutron entanglement up to \NNLO}
We now proceed to analyze the entanglement power at higher orders in WPC, up to \NNLO{}, for both $np$ and $nn$ systems, using potentials from Ref.~\cite{Carlsson:2015vda}.

\begin{figure}[ht]
    \centering
    \includegraphics[width=\linewidth]{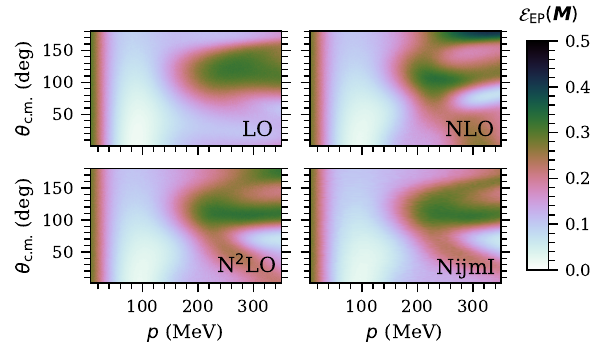}
    \caption{Entanglement power of the $M$ matrix $\mathcal{E}_{\text{EP}}(\boldsymbol{M})$ for the $np$ system using the WPC potential at different chiral orders and using the phenomenological NijmI potential~\cite{Stoks:1994wp}. The LECs in the WPC potentials are from Ref. \cite{Carlsson:2015vda}. $\mathcal{E}_{\text{EP}}(\boldsymbol{M})$ is evaluated for scattering angles $\theta_{\text{c.m.}}\in[1^\circ, 180^\circ]$ and relative momenta $p\in[7,350]$ MeV.}
    \label{fig:EP_LO_NLO_N2LO_Nijm_np}
\end{figure}

The resulting order-by-order entanglement pattern for $np$ scattering is shown in Fig.~\ref{fig:EP_LO_NLO_N2LO_Nijm_np}. The phenomenological result, using the NijmI potential~\cite{Stoks:1994wp}, is included in the lower right panel for comparison, with data collected from NN-OnLine~\cite{nn-on-line}. As before, we present results for the nonforward scattering directions. 

There is a clear convergence towards the empirical result and the N$^2$LO potential gives an accurate description of the entanglement power in the considered kinematical region. 
Angular complexity increases for $p \gtrsim 200$~MeV already at NLO, where two-pion exchange enters in WPC.  

The persistent suppression of spin entanglement observed for $50\lesssim p\lesssim 150$ MeV indicates that the Wigner $SU(4)$ symmetry is approximately a good symmetry in this energy domain. Long-range, symmetry-breaking terms of the nuclear interaction, such as the tensor force from OPE, induce entanglement at slightly higher energies. The low-energy entanglement, for momenta $p< 1/a_S$, is generated by the short-distance symmetry-breaking terms in the contact potential. 

\begin{figure}[ht]
    \centering
    \includegraphics[width=\linewidth]{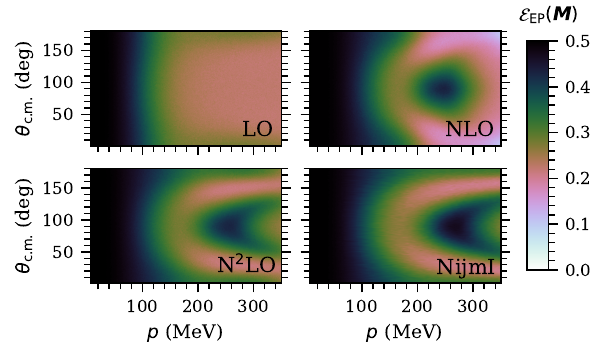}
    \caption{Entanglement power of the $M$ matrix $\mathcal{E}_{\text{EP}}(\boldsymbol{M})$ for the $nn$ system using the WPC potential at different chiral orders. The lower right panel is the $pp$ result using the NijmI potential~\cite{Stoks:1994wp} without the electromagnetic contribution. Due to charge symmetry, the nuclear $pp$ and $nn$ M matrices should be very similar. The LECs in the WPC potentials are from Ref. \cite{Carlsson:2015vda}. $\mathcal{E}_{\text{EP}}(\boldsymbol{M})$ is evaluated on the same angle-energy grid as in \cref{fig:EP_LO_NLO_N2LO_Nijm_np}.}
    \label{fig:EP_LO_NLO_N2LO_nn_Nijm_pp}
\end{figure}

We now turn to $nn$ scattering. \Cref{fig:EP_LO_NLO_N2LO_nn_Nijm_pp} presents the entanglement power in the WPC framework, order by order up to \NNLO. 
For reference, the lower right panel shows the entanglement power of the nuclear $pp$ NijmI $M$ matrix from NN-OnLine~\cite{nn-on-line} (which does not include $nn$ scattering data). 
Due to charge symmetry, this result is expected to closely approximate the $nn$ case.

The entanglement patterns are very different for $np$ and $nn$ scattering, as seen when comparing Figs.~\ref{fig:EP_LO_NLO_N2LO_Nijm_np} and \ref{fig:EP_LO_NLO_N2LO_nn_Nijm_pp}. In general, the entanglement power is much stronger for $nn$ scattering where the Pauli principle for indistinguishable particles limits the partial waves that contribute to the computation of $M$-matrix elements. 
More specifically the entanglement suppression seen for the $np$ system in the region $50 \lesssim p \lesssim 150$~MeV is replaced by very large entanglement in the $nn$ case. 

At low energies where $S$-wave scattering dominates, the $nn$ $M$ matrix becomes proportional to an operator that projects the initial spin state to the maximally entangled spin-singlet state due to the Pauli principle, see \cref{eq:very_low_energy_M_matrix}. This has already been noted in Ref. \cite{Hu:2025lua}. The absence of entanglement suppression in the $nn$ system can be attributed to the inapplicability of Wigner $SU(4)$ symmetry in this case. For even-parity waves, Wigner $SU(4)$ symmetry would imply identical phase shifts for all partial waves with the same orbital angular momentum $L$. However, in the $nn$ system, only spin-singlet channels exist for even $L$, making such a condition ill-defined. It is also worth reiterating that Serber symmetry is trivial in the $nn$ case due to the identical nature of the two nucleons.

\subsection{Entanglement power in the forward scattering direction}
For completeness, we now extend the analysis of the previous sections by also studying entanglement power in the forward scattering direction. In particular, we will relate the entanglement power at $\theta_{\text{c.m.}}=0^\circ$ calculated with the full $S$ matrix to the pure $S$-wave formula derived in Ref. \cite{Beane:2018oxh}.

\begin{figure}[ht]
    \centering
    \includegraphics[width=\linewidth]{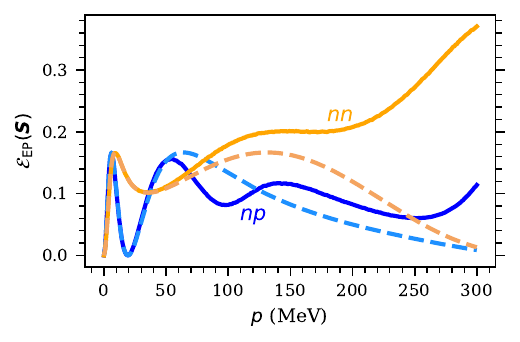}
    \caption{Entanglement power of the $S$ matrix in the forward scattering direction using the N$^2$LO WPC potential for $np$ and $nn$ scattering. The solid lines are the result for the full $S$ matrix while the dashed lines are the $S$-wave result using the analytical formula for the entanglement power as derived in Ref. \cite{Beane:2018oxh}, see Eqs. \eqref{eq:S_wave_EP_np} and \eqref{eq:S_wave_EP_nn}. Note that the exact solution and the $S$-wave solution align for momenta $p\lesssim 50$ MeV for both $np$ and $nn$ scattering.}
    \label{fig:forward_scattering}
\end{figure}

For an incoming plane wave state, the trivial part of the scattering operator only contributes in the forward scattering direction such that the outgoing spin state $\ket{\chi_{\text{out}}}$ is given by
\begin{equation}
    \ket{\chi_{\text{out}}} \propto (\mathbbm{1} + 2ip\boldsymbol{M}(\boldsymbol{p},\boldsymbol{p}))\ket{\chi_{\text{in}}} \equiv\boldsymbol{S}\ket{\chi_{\text{in}}},
    \label{eq:forward_direction_main_text}
\end{equation}
see also \cref{app:forward_direction}. At low-energies, the $M$ matrix in \cref{eq:forward_direction_main_text} will only get contributions from $S$-waves, and the entanglement power can be expressed analytically
\cite{Beane:2018oxh,Bai:2023tey}
as
\begin{equation}
    \mathcal{E}_{\text{EP}}^{np}(\boldsymbol{S}) = \frac{1}{6}\sin^2\left( 2(\delta_{{^3\!S_1}}-\delta_{{^1\!S_0}})\right),
    \label{eq:S_wave_EP_np}
\end{equation}
and
\begin{equation}
    \mathcal{E}_{\text{EP}}^{nn}(\boldsymbol{S}) = \frac{1}{6}\sin^2(2\delta_{{^1\!S_0}}).
    \label{eq:S_wave_EP_nn}
\end{equation}
The entanglement power of the $S$ matrix in the forward direction is shown in \cref{fig:forward_scattering} for both $np$ and $nn$ scattering using the N$^2$LO WPC potential from Ref.~\cite{Carlsson:2015vda}. The predictions from the analytical $S$-wave formulas, \eqref{eq:S_wave_EP_np} and \eqref{eq:S_wave_EP_nn}, are also displayed. The exact results and the $S$-wave approximation agree well at momenta $p\lesssim 50$ MeV.

Note that the entanglement power for the $S$ matrix is discontinuous at $\theta_{\text{c.m.}}=0^\circ$. The discontinuity is most pronounced at very low energies where the entanglement power vanishes in the forward scattering direction (for both $np$ and $nn$ scattering) due to the dominance of the trivial component of the $S$ matrix (see \cref{eq:forward_direction_main_text}). In contrast, the entanglement power remains nonzero at low energies in non-forward directions for both $np$ and $nn$ scattering (see \cref{fig:EP_LO_NLO_N2LO_Nijm_np,fig:EP_LO_NLO_N2LO_nn_Nijm_pp}).

The trivial part of the $S$ matrix can also be exposed---without making an explicit assumption about the scattering direction---by letting the initial state be an isotropic $S$ wave. Then, the angular delta function in \cref{eq:scattering_operator} vanishes which leaves the identity operator, see \cref{app:S_waves}.

The zero entanglement power for $np$ scattering around $p\approx 20$ MeV in \cref{fig:forward_scattering} is due to the fact that the difference between the ${^1\!S_0}$ and ${^3\!S_1}$ phase shifts is $\pi/2$, see Eq. \eqref{eq:S_wave_EP_np} and the discussion in Ref.~\cite{Beane:2018oxh}. This is not a manifestation of Wigner $SU(4)$ symmetry because the phase shifts are different. A similar zero point does not appear for $nn$ scattering where there is a single $S$-wave phase shift. Finally, we note that the absence (presence) of entanglement suppression for $nn$ ($np$) scattering is also visible in the forward direction for $p\gtrsim50$ MeV. 

\section{Conclusions and Outlook}
\label{sec:conclusion}
In this work, we have explored the connection between accidental nuclear symmetries and spin entanglement for $np$ and $nn$ scattering in the context of $\chi$EFT. We systematically embedded enhanced Wigner $SU(4)$ and Serber symmetries into multiple variants of LO $\chi$EFT potentials and employed the full scattering matrix to perform a detailed and quantitative analysis of their impact on pure spin entanglement in $np$ scattering. Furthermore, we studied spin entanglement up to \NNLO{} in WPC for both $np$ and $nn$ scattering. Finally, we highlighted the distinct behavior between the entanglement power in forward and nonforward scattering. We also established connections to earlier studies that were restricted to $S$-wave contributions, thereby extending the scope of entanglement analysis beyond previously considered regimes.

We conclude that Wigner $SU(4)$ and Serber symmetry breaking in different components of the LO potential generate spin entanglement in specific kinematical regions for $np$ scattering. This shows that entanglement can be used as a quantitative measure for investigating accidental symmetries in nuclear interactions. In particular, we find that Wigner $SU(4)$ symmetry breaking in the contact terms leads to a pronounced enhancement of entanglement power at low energies. In contrast, the $SU(4)$-breaking tensor force predominantly drives spin entanglement at higher momenta ($p\gtrsim150$ MeV) and near-backwards scattering ($100^\circ\lesssim\theta_{\text{c.m.}}\lesssim150^\circ$), as revealed through a full $S$-matrix analysis.
It does remain to study how robust the observed entanglement suppression is when taking EFT truncation errors into account.

A striking feature in the $np$ system is the pronounced suppression of entanglement in the intermediate energy region $50 \lesssim p \lesssim150$~MeV. This valley of low entanglement disappears when considering $nn$ scattering, in which higher spin entanglement is observed in the entire kinematical domain. This observation can be attributed to the presence (absence) of Wigner $SU(4)$ symmetry in the $np$ ($nn$) system. The significant low-energy entanglement observed in the $nn$ case is due to the Pauli principle, which constrains the low-energy $nn$ $S$ matrix to be proportional to a Bell projector. 

In the special case of forward scattering we observe the absence (presence) of entanglement suppression in the $nn$ ($np$) system for $p\gtrsim50$~MeV, precisely as for the nonforward directions. Moreover, the entanglement power is necessarily zero as $p\rightarrow0$ due to the dominance of the trivial component of the scattering operator for forward scattering.

This work has shown that entanglement provides a new perspective on the role of accidental symmetries and the relative importance of different terms in the nuclear force. 
We therefore advocate for future studies on entanglement-guided power-counting schemes in nuclear effective field theory to explore how accidental symmetries could potentially be incorporated in leading and (perturbative) subleading orders of a realistic nuclear interaction. 
A natural extension is to explore approximate Wigner $SU(4)$ symmetry at higher orders. For example, the two-pion exchange part of the NLO potential is of the form 
\begin{equation}
    W_C(\boldsymbol{q})\boldsymbol{\tau}_1\cdot\boldsymbol{\tau}_2 + V_S(\boldsymbol{q})\boldsymbol{\sigma}_1\cdot\boldsymbol{\sigma}_2 + V_T(\boldsymbol{q})\boldsymbol{\sigma}_1\cdot\boldsymbol{q}\boldsymbol{\sigma}_2\cdot\boldsymbol{q},
\end{equation}
where the last term can be split into one piece proportional to $\boldsymbol{\sigma}_1\cdot\boldsymbol{\sigma}_2$ and another proportional to the tensor operator. The situation is more complex than at LO due to the more complicated function $V_T(\boldsymbol{q})$. One might have to numerically split this term into Wigner $SU(4)$ symmetric and breaking parts.

As a final outlook, we mention the relevance of our findings for understanding entanglement entropies in many-nucleon systems~\cite{Robin:2020aeh, Kruppa:2021yqs, Tichai:2022bxr, Johnson:2022mzk}. Although we have focused on the entanglement power of nucleon-nucleon scattering---where the initial state corresponds to spatially separated particles---we speculate that the observed entanglement suppression and its connection to accidental symmetries can explain the weak entanglement between neutron- and proton species observed in many-body systems~\cite{Johnson:2022mzk,Perez-Obiol:2023wdz} and provide further motivation for the weak-entanglement approximation for nuclear structure~\cite{Gorton:2024hbb}.

\begin{acknowledgments}
A.\ L.\ C.\ would like to thank Tanja Kirchner for helpful discussions on entanglement and identical particles. The authors are grateful for Wigner $SU(4)$ discussions with Simone Salvatore Li Muli. C.F.\ specifically acknowledges discussions with Daniel Phillips and Simone Salvatore Li Muli on Wigner $SU(4)$ symmetry in $\chi$EFT. This work was supported by the Swedish Research Council (Grants No.~2020-05127 and No.~2021-04507). 
\end{acknowledgments}
\FloatBarrier

\appendix
\section{}
\label{appendix}
In this appendix, we derive the expression for the outgoing spin state in the forward scattering direction. We also calculate the outgoing spin state when the initial state is assumed to be an $S$ wave and the $S$ matrix is truncated at $L=L'=0$. 

\subsection{Forward direction}
\label{app:forward_direction}
Let the initial state be $\ket{\psi_{\text{in}}} = \ket{\boldsymbol{p}}\ket{\chi_{\text{in}}}$. The outgoing state $\ket{\psi_{\text{out}}}$ is then given by 
\begin{equation}
    \begin{split}
        \ket{\psi_{\text{out}}} =  &S\ket{\boldsymbol{p}}\ket{\chi_{\text{in}}}\\
        =&\int d\Omega_{p'}\ket{\boldsymbol{p}'}\Big(\delta^{(2)}(\hat{\boldsymbol{p}}'-\hat{\boldsymbol{p}})\mathbbm{1}\\
        &+\frac{ip}{2\pi}\boldsymbol{M}(\boldsymbol{p}', \boldsymbol{p})\Big)\ket{\chi_{\text{in}}} \quad (p'=p).
    \end{split}
\end{equation} 
The overlap of $\ket{\psi_{\text{out}}}$ with a general wave packet $\ket{\varphi} = \int d^3k\,\varphi(\boldsymbol{k})\ket{\boldsymbol{k}}$ is
\begin{equation}
    \begin{split}
        \braket{\varphi|\psi_{\text{out}}} = &\int d\Omega_{p'}\,\varphi^*(\boldsymbol{p}')\Big(\delta^{(2)}(\hat{\boldsymbol{p}}'-\hat{\boldsymbol{p}})\mathbbm{1}\\
        &+\frac{ip}{2\pi}\boldsymbol{M}(\boldsymbol{p}', \boldsymbol{p})\Big)\ket{\chi_{\text{in}}}\quad {(p'=p)}.
    \end{split}
\end{equation}
For scattering in the forward direction, we assume that the wave packet is peaked around $\boldsymbol{p}'=\boldsymbol{p}$. Then, $\boldsymbol{M}(\boldsymbol{p}',\boldsymbol{p}) \approx \boldsymbol{M}(\boldsymbol{p},\boldsymbol{p})$ and $\varphi^*(\boldsymbol{p}')\approx \varphi^*(\boldsymbol{p})$, so that
\begin{equation}
    \begin{split}
        \braket{\varphi|\psi_{\text{out}}} 
        &=\varphi^*(\boldsymbol{p})(\mathbbm{1} +2ip\boldsymbol{M}(\boldsymbol{p},\boldsymbol{p}))\ket{\chi_{\text{in}}}.
    \end{split}
\end{equation}
Thus, the outgoing spin state $\ket{\chi_\text{out}}$ is given by
\begin{equation}
    \ket{\chi_{\text{out}}} \propto  (\mathbbm{1} + 2ip\boldsymbol{M}(\boldsymbol{p}, \boldsymbol{p}))\ket{\chi_{\text{in}}} \equiv \boldsymbol{S}\ket{\chi_{\text{in}}} ,
    \label{eq:forward_direction}
\end{equation}
in the forward scattering direction.

\subsection{S waves}
\label{app:S_waves}
We now derive the expression for the outgoing spin state when the initial state is assumed to be an $S$-wave state. This results in the same $S$-wave $S$ matrix as given in e.g., Refs. \cite{Beane:2018oxh, Low:2021ufv}. 

Let the initial state be denoted by $\ket{\psi_{\text{in}}} = \ket{pLM_L}\ket{\chi_{\text{in}}}$. The following convention is used for the overlap between the plane wave and the spherical bases
\begin{equation}
    \braket{\boldsymbol{k}|pLM_L} = i^{-L}Y_{LM_L}(\hat{\boldsymbol{k}})\frac{\delta(k-p)}{p^2}.
\end{equation}
The out state $\ket{\psi_{\text{out}}}$ is then given by 
\begin{equation}
    \begin{split}
        \ket{\psi_{\text{out}}} = &\int d\Omega_{p'}\int d^3p''\ket{\boldsymbol{p}'}\Big(\delta^{(2)}(\hat{\boldsymbol{p}}''-\hat{\boldsymbol{p}}')\mathbbm{1}\\
        &+\frac{ip''}{2\pi}\boldsymbol{M}(\boldsymbol{p}',\boldsymbol{p}'')\Big)\braket{\boldsymbol{p}''|pLM_L}\ket{\chi_{\text{in}}}\quad (p'=p'').\\
    \end{split}
\end{equation}
For an initial $S$-wave state ($Y_{00}(\hat{\boldsymbol{k}}) = \sqrt{1/4\pi}$) 
\begin{equation}
    \begin{split}
        \ket{\psi_{\text{out}}} = &\int d\Omega_{p'}\int d\Omega_{p''}\ket{\boldsymbol{p}'}\Big(\delta^{(2)}(\hat{\boldsymbol{p}}''-\hat{\boldsymbol{p}}')\mathbbm{1}\\
        &+\frac{ip}{2\pi}\boldsymbol{M}(\boldsymbol{p}',\boldsymbol{p}'')\Big)\sqrt{\frac{1}{4\pi}}\ket{\chi_{\text{in}}}\quad (p'=p''=p).
    \end{split}
\end{equation}
In the $S$-wave approximation, $M(\boldsymbol{p}', \boldsymbol{p}'') = M(p)$, and the overlap between $\ket{\psi_{\text{out}}}$ and a general wave packet $\ket{\varphi}$ is 
\begin{equation}
    \begin{split}
        \braket{\varphi|\psi_{\text{out}}} &=(\mathbbm{1} + 2ip\boldsymbol{M}(p))\ket{\chi_{\text{in}}}\sqrt{\frac{1}{4\pi}}\\
        &\times \int d\Omega_{p'}\varphi^*(\boldsymbol{p}') \quad (p'=p).
    \end{split}
\end{equation}
Thus, the outgoing spin state for the $S$-wave approximated $S$ matrix and an incoming $S$-wave state is given by
\begin{equation}
    \ket{\chi_{\text{out}}} \propto (\mathbbm{1} + 2ip\boldsymbol{M}(p))\ket{\chi_{\text{in}}}.
    \label{eq:chi_out_initial_S_wave_state}
\end{equation}
In Eq. \eqref{eq:chi_out_initial_S_wave_state}, we note that there is no dependence on $\theta_{\text{c.m.}}$, because $S$ waves are isotropic, and that the trivial part of the scattering operator is included.

\bibliography{bib} 

\end{document}